\documentclass[12pt]{article}
\usepackage{epsfig,graphicx}

\begin{document}
\title{Skating on Slippery Ice}

\author{J.M.J. van Leeuwen}
\maketitle
\begin{center}
Instituut-Lorentz, Universiteit Leiden,\\
Niels Bohrweg 2, 2333 CA Leiden, The Netherlands.
\end{center}

\begin{abstract}
The friction of a stationary moving skate on smooth ice is investigated,
in particular in relation to the formation of a thin layer of water between 
skate and ice. It is found that the combination of ploughing and melting 
gives a friction force that is rather insensitive for parameters 
such as velocity and temperature. The weak dependence originates from
the pressure adjustment inside the water layer. For instance, higher velocities,
giving rise to higher friction, also lead to larger pressures, 
which, in turn, decrease the contact zone and so lower the friction.
By treating ice as a Bingham solid the theory combines and completes two
existing but conflicting theories on the formation of the water layer. 
\end{abstract}

keywords: solid friction, fluid mechanics, lubrication.

\section{Introduction}

Ice seems to be the only substance on which one can conveniently skate,
which prompts the question: ``what sort of special properties does ice have 
as compared to other solids?'' Moreover one can glide on ice over a wide
range of velocities, types of skates and temperatures. Ice is in many respects
a peculiar solid and there is much folklore about the mystery of skating.

Ice is one of the few substances where the solid is less dense than the
liquid, which has a profound impact in nature.  Skating is a minor beneficiary 
of this property, as canals freeze on top, so one does not have to wait till 
the canal is solidly frozen. 
Another interesting property of water is that the melting line in the 
pressure-temperature plane has an unusual slope: with increasing pressure 
the freezing temperature lowers, while usually pressure favours the solid phase.
It is illustrated in the famous high-school experiment where
a steel cable with weights on both sides, melts itself through a block of ice
at temperatures below zero, such that the block refreezes on top of the steel 
cable! This property has featured for quite a while as explanation for 
skating: due to the pressure exerted by the skater on the ice, a water layer 
forms and the skates glide on this water layer. It has been demonstrated 
several times that this explanation is not feasible \cite{bowden,schenau}.
Although the lowering of the melting point under pressure does not 
explain the skating phenomenon, its influence can not be dismissed 
at low temperatures, as we will show. 

The slipperiness of ice has also been attributed to the special structure
of the free ice surface. The existence of a water layer on the surface, 
even without skating, was already suggested by Faraday \cite{faraday}. 
Computations and measurements indicate that this layer is only a few molecules
thick, such that one cannot speak of this water layer as a hydrodynamic system,
see e.g. \cite{rosenberg}.
For slow velocities and low temperatures the structure of the
surface plays an important role on the friction properties \cite{amsterdam}.

In this paper we study the formation and influence of the water layer 
underneath the skate for usual conditions, i.e for sliding velocities of 
meters per second and temperatures of a few degrees below the melting point
of ice. Gliding is only a part of the physics of skating. 
Also important is the ability to push oneself forward, which is possible due
to the shape of the skate and to the fact that ice is easily deformable. 

The main argument for the formation of a water layer, is
that friction generates heat and that heat melts ice. How much of the heat
melts ice and which part leaks away, is an important issue, which we address
in this paper. We will treat
the water layer as a hydrodynamic system, which implies that its thickness
has to be at least of the order of 10 nm. If such a layer of water is formed,
the hydrodynamic properties of the layer determine the friction, which then
becomes independent of the surface properties.

The physics of the water layer between skate and ice is not simple, with
a rich history, see e.g. \cite{schenau,rosenberg,persson}.
In spite of the fact that the problem is century old, the water layer has 
never been directly observed.  A potential method for observation is based 
on the difference in dielectric properties of ice and water at high 
frequencies \cite{leiden}. Indirect evidence for the water layer may
result from measurements of the friction of a skate on ice. 
If friction is mediated through a water layer, then its
characteristics can be checked. This paper deals with a calculation
of the friction.

It is well known that a skater on virgin ice leaves a trail. Is this trail
due to melting or to plastic deformation (ploughing)? The deformation is
plastic if the exerted pressure exceeds the hardness of ice. The trail is an
indication that the deformation of ice is plastic. Indeed, the weight of a
skater of, say 72 kg, cannot be supported by an elastic deformation of
ice. Moreover skates have sharp edges which will make kinks in the
surface of ice (even in horizontal position) and near a kink the 
pressure will always exceed the 
hardness of ice. Therefore we focus on plastic deformation of the ice and
justify this {\it a posteriori} by the high pressures occurring in the
water layer for skating speeds. 

At the moment there are two quantitative but competing theories for the 
formation of a water layer and the furrow of the trail.
The one by Lozowski and Szilder \cite{lozowski}, assumes that most of the dent
in the ice is the result of ploughing. The other theory, by
Le Berre and Pomeau \cite{pomeau} assumes that the dent  is due
to melting only. We will show which fraction of the trail is 
due to melting and which is due to ploughing. 
The two regimes, melting and ploughing merge continuously. Although our 
description is a unification of both theories, the results are substantially 
different from both theories. 

In this paper we discuss the issue in the
simplest possible setting: a speed skater moving in upright position
over the ice with a velocity $V$ on perfectly smooth ice and skates. 
The skater stands with his mass $M$ on
one skate. For skating near the melting point of ice, heat flows into the 
skates and into the ice are less important and we discuss their influence 
later on. 
Our main concern is the thickness of the water layer {\it underneath} the skate;
the water films that form at the {\it sides} of the skate,  play a minor role.

The only measurements of the friction of skates under realistic 
conditions, that we are aware of, have been performed by de Koning et al. 
\cite{schenau}.
Their skater had a velocity of speed of $V=8$ m/s and a weight of 72 kg.
Together with the standard parameters of skates:
curvature $R=22$ m and width $w=1.1$ mm, we call these specifications the 
skating conditions. Unless otherwise stated, our calculations are carried 
out for temperature $T=0 ^0$C. We take the skating conditions as 
reference point and
vary the parameters individually with respect to this point.

The various aspects of the theory are presented in Sections in the following 
order:
\begin{description}
\item[\ref{geo}]  describes the used coordinate systems and
the geometry of the skate.
\item[\ref{hardness}] provides the necessary information on the material
constants of water and ice.
\item[\ref{static}] gives the force balance for a static skater.
\item [\ref{heat}] derives the heat balance, determining 
the thickness of the water layer.
\item[\ref{melting}] solves the equations for the thickness of the water layer
in the regime where only melting plays a role.
\item[\ref{hydro}] summarises the necessary formulas for hydrodynamics 
and pressure of the water layer.
\item[\ref{ploughing}] yields the shape of the water layer in the 
ploughing regime.
\item[\ref{cross}] treats the cross-over from the ploughing to the 
melting regime.
\item[\ref{pressure}] calculates the pressure in both regimes.
\item[\ref{forces}] relates the weight of the skater to the pressure in the
water layer.
Also the slowing down force of the ice is computed, 
which is the sum of the friction in the water layer and the ploughing force.
\item[\ref{velocity}] contains the velocity dependence of the friction.
\item[\ref{temper}] discusses the influence of the ice temperature on the
friction.
\item[\ref{discussion}] closes the paper with a discussion of the 
approximations and a comparison with the existing theories.
\end{description}
In addition a number of separate issues are treated in Appendices.

\section{Geometry of the skates}\label{geo}

\begin{figure}[h]
\vspace*{-2cm}

\begin{center}
  \epsfxsize=0.7\linewidth  \epsffile{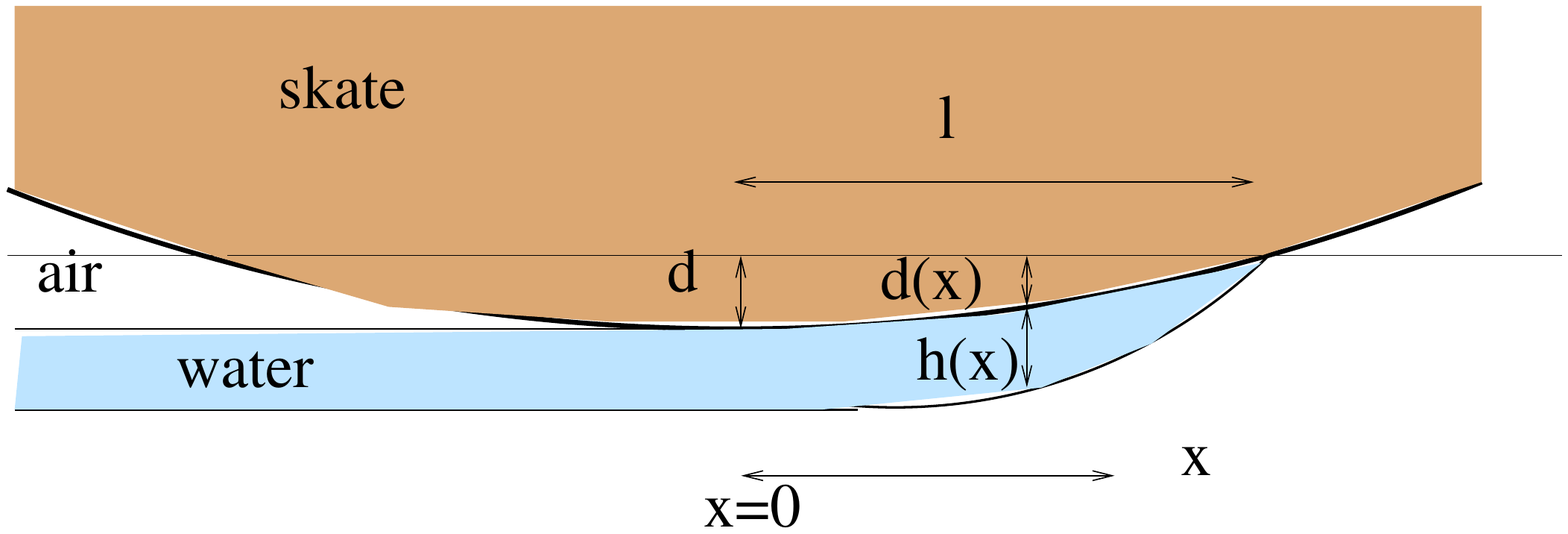}
    \vspace*{-2cm}

  \caption{longitudinal cross-section of skate, ice and water layer in between.}
\label{long}
\end{center}
\end{figure}

For the description of the phenomena we need two coordinate systems:
the ice fixed system and that of the moving skater. If $x,y,z$ are
the coordinates in the ice system, then the coordinates 
$x',y',z'$ of the same point in the skate system are
\begin{equation} \label{a1}
x'= x - V t, \quad \quad y'=y, \quad \quad z'=z,
\end{equation} 
where V is the velocity of the skate. The $x$ coordinate points in the
forward direction of the skate. The origin of the skate coordinates is
in the middle of the skate at the level of the ice. The lowest point of 
the skate, the depth of the trail, is a distance $d$ below the original 
ice level.
The $y$ direction is horizontally and perpendicular to the skate blade
and the $z$ direction points downward into the ice. At time $t=0$ the
two coordinate systems coincide. See Fig.~\ref{long} for a cross-section
in the longitudinal direction.
$d(x')$ is the locus of the bottom of the skate. With $R$ the curvature
radius of the skate it is given by the equation
\begin{equation} \label{a2}
[R-d +d(x')]^2 + x'^2 = R^2, \quad \quad {\rm or} \quad \quad 
d(x') = [R^2 -x'^2]^{1/2} + d -R.
\end{equation} 
In the ice system we have correspondingly
\begin{equation} \label{a3}
d(x,t)=d(x')=[R^2 -(x-Vt)^2]^{1/2} + d -R.
\end{equation} 
So for a fixed point $x$ in the ice system, the downward velocity of the skate 
$v_{\rm sk} (x)$ is at $t=0$
\begin{equation} \label{a4}
v_{\rm sk} (x) = \left(\frac{\partial d(x,t)}{\partial t} \right)_{t=0} 
= V \frac{x}{[R^2 -x^2]^{1/2}} \simeq V\frac{x}{R}.
\end{equation}
The last approximation uses that $x$ is a few centimeters and $R$ about 
20 meters. $v_{\rm sk}$ is also the velocity with
which the top of the water layer, in contact with the skate, comes down. 
Later on we need also  $v_{\rm ice} (x)$, being the velocity at the bottom 
of the layer with which the ice recedes due to the pressure.
\begin{figure}[h]
\vspace*{-2cm}

\begin{center}
  \epsfxsize=0.7\linewidth  \epsffile{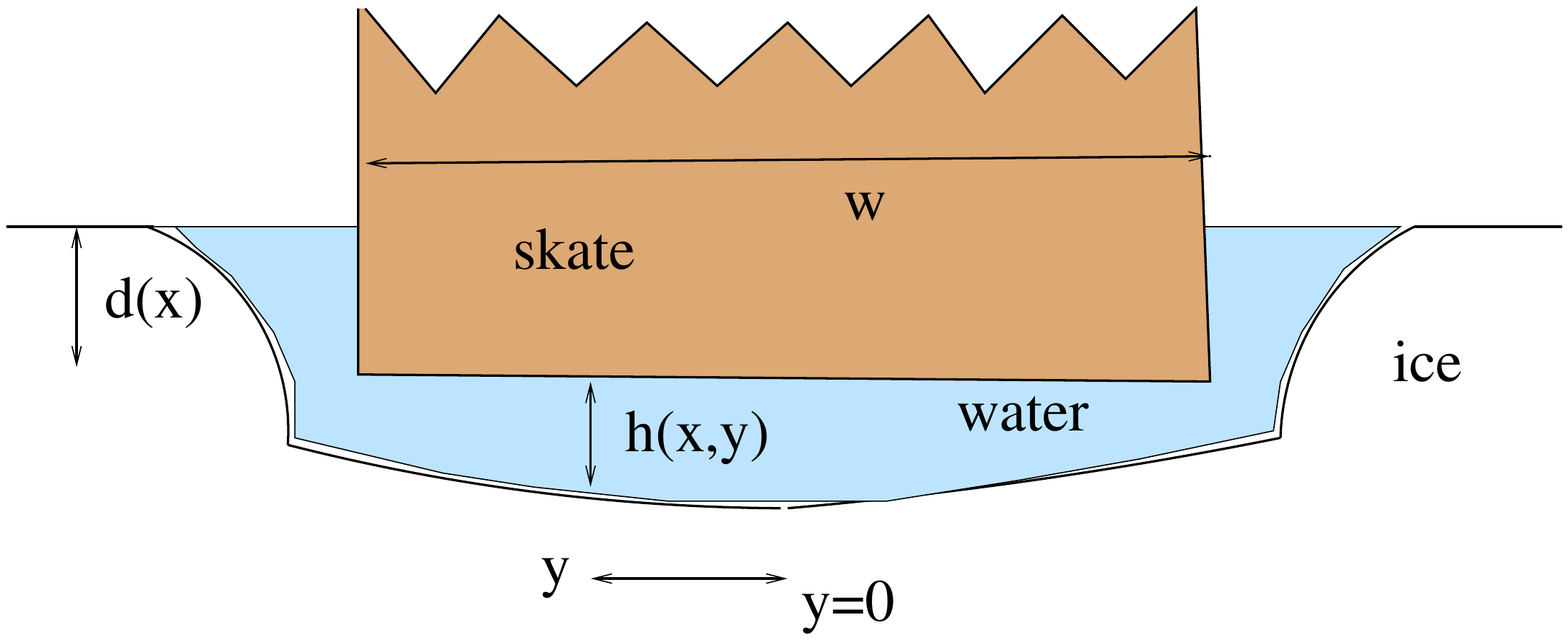}
    \vspace*{-1cm}

  \caption{transverse cross-section in the $y,z$ plane of the skate and 
the layer of water underneath. Note that, for visibility, width (mm) and depth 
($\mu$m) are not drawn in proper scale.}
\label{trans}
\end{center}
\end{figure}

The thickness of the water layer at a point $x$ is denoted by $h(x,y)$. So in
the downward direction we have the skate between $0<z<d(x)$, water between 
$d(x)<z<d(x)+h(x,y)$ and ice below $z>d(x)+h(x,y)$. The water at
the sides of the skate is of minor influence, since the depth $d(x)$ 
measures in $\mu$m, while the width of the skate is around 1 mm. In order
to focus on the essentials we restrict the discussion to the treatment of
the layer underneath the skate.
In Fig.~\ref{trans} we give a sketch
of the transverse cross-section in the $y,z$ plane. As indicated in this
figure, the water layer may vary in the transverse $y$ direction. In the 
coming sections we approximate $h(x,y)$ by a function $h(x)$ of $x$ alone.
In Appendix \ref{ydep}, we show that this is a good approximation for 
calculating the friction.

\section{Material constants of water and ice}\label{hardness}
\begin{table}
\begin{center}
\begin{tabular}{||l|l|l|c||}
\hline
 & & &\\*[-2mm]
material constant &symbol & value & unit \\*[2mm]
\hline
 & & &\\*[-2mm]
dynamic viscosity water & $\eta$ & 1.737*10$^{-3}$ & Pa s \\*[2mm]
thermal conductivity water &  $\kappa_{\rm w}$  & 0.591 & J/(m s K) \\*[2mm]
thermal conductivity ice & $\kappa_{\rm ice}$ & 1.6 & J/(m s K) \\*[2mm]
thermal diffusivity ice & $\alpha_{\rm ice}$ & 0.843*10$^{-6}$ & m$^2$/s \\*[2mm]
density water and ice & $\rho$  & $10^3$ & kg/m$^3$ \\*[2mm]
latent heat of melting & $\rho L_H $ & 0.334*10$^9$ & J/m$^3$ \\*[2mm]
Young's modulus ice & $E$ & 0.88*$10^9$ & Pa \\*[2mm]
\hline 
\end{tabular}
\caption{material constants of water and ice}\label{mater}
\end{center}
\end{table}
In the Table \ref{mater}  we have listed the relevant material constants 
of water and ice. Apart from these well known constants, there
are two more material properties relevant for skating: the hardness
of ice $p_H$ and the deformation rate $\gamma$.
The Brinell hardness number is measured by pushing with a force $F$, 
an ``undeformable'' spherical ball into the material. 
After lifting the force, the material shows a dent, with surface $S$. 
The ratio $F/S$ is independent of $F$ and equal to the hardness $p_H$. 
This means that the material reacts upon 
deforming forces with a fixed counter pressure $p_H$, such that the contact 
surface $S$ times $p_H$ balances the applied force $F$.\footnote{The Brinell
hardness takes as contact surface the spherical surface of the dent, which 
is slightly larger than the top circle of the dent. In contrast to the 
Brinell hardness, we measure the contact area in the horizontal direction 
and not along the skate, since the horizontal surface matters
for the force balance Eq.~(\ref{a2}).}

For the hardness dependence of ice on the temperature Pourier et al. 
\cite{pourier} give the relation
\begin{equation} \label{a5}
p_H = (14.7 - 0.6\, T)*10^{6} \, {\rm Pa},
\end{equation} 
with $T$ the temperature in centigrades. An earlier measurement gave quite
different values \cite{new}. The value depends on the method of measurement 
\cite{amsterdam}. We take the viewpoint that the hardness is defined by
the response to a quasi-static deformation of the ice. 
Mostly the hardness comes into our analysis
as a multiplicative constant. Although the measurements of Pourier et al.
were not carried out quasi-statically, we stick to the value given in 
Eq.~(\ref{a5}) for the hardness in our calculations, when explicitly needed.

However, skating is a dynamic event. For instance a forward skating 
velocity of 10 m/s implies, a downward velocity of about 1 cm/s 
at the tip of contact. In order that the ice recedes at such a large rate, 
one needs pressures far exceeding the hardness. Such large pressures
require a relation between the applied pressure and the velocity with 
which the ice recedes.
With $p(x,y,d(x)+h(x))$ the
pressure in the water layer in contact with the ice, 
we will use for the downward velocity of ice the relation
\begin{equation} \label{a8}
v_{\rm ice} (x,y) = \gamma [p(x,y,d(x)+h(x))-p_H],
\end{equation}    
where $\gamma$ is a material constant with the dimension [m/(Pa s)].
Eq.~(\ref{a8}) takes the receding velocity proportional to the pressure excess.
This is similar to treating ice as a Bingham solid \cite{nye}, where one
puts, for plastic flow, the shear rate proportional to pressure excess.
The deformed region of the ice is of the order of the width $w$. So dividing 
$v_{\rm ice}$ by $w$ gives the order of the occurring shear rates. In this way
we deduce, from the measured shear rates \cite{barnes,karna}, a value 
$\gamma p_H \simeq 1$ mm/s. This is not more than an order of magnitude
estimate, since glaciers and laboratory experiments induce plastic 
flows on a time scale much lower than in speed skating.
 
\section{Static deformation} \label{static}

Elastic deformations of ice are controlled by the elastic
coefficient (Youngs modulus). 
By calculating the elastic deformation field due to a skate which bears a 
weight $M$, one estimates that for $M$ below 10kg, the skate makes an 
elastic deformation. The estimate is hampered by role of the edges of the
skate. If they are not rounded off a bit, they produce a kink in the
deformation field, which leads to unlimited pressures in the ice. The estimate
shows, however, that for practical skater masses the deformation is plastic. 

Static inelastic deformations are determined by the hardness $p_H$.  
At rest, the skater exerts a pressure on the ice equal to the hardness $p_H$. 
The contact area times the pressure balances the weight of the skater. 
The contact area is the width $w$ of the skate times the contact length $2l_0$.
So one has the force balance
\begin{equation} \label{a6}
M g = 2 p_H w l_0,
\end{equation} 
which gives the value of $l_0$. The static depth $d_0$ of the dent in 
the middle of the skate is related to $l_0$ by geometry 
\begin{equation} \label{a7}
R^2 = (l_0)^2 + (R-d_0)^2, \quad \quad \quad {\rm or} \quad \quad \quad 
d_0 \approx \frac{l_0^2}{2 R}.
\end{equation}
The two equations (\ref{a6}) and (\ref{a7}) determine the static values 
of $l_0$ and $d_0$. We find for a weight of 72 kg the values 
$l_0 = 2.2$cm and $d_0 = 11 \mu$m. We note that this estimate assumes that
the pressure distribution in the ice underneath the skate is uniformly 
equal to $p_H$. If one calculates, for small weights, the pressure 
distribution for elastic deformations, one finds that the pressure is largest
at the edges of the skate and in the middle where the deformation
is deepest. Thus at the point where the elastic deformation turns gradually
into a plastic deformation the above estimate does not apply. It only
applies for a fully developed plastic deformation. 

The calculation of contact length $l$ and the depth $d$ for a moving skater 
is a major part of the problem. The relation between $l$ and $d$ is the same
as Eq.~(\ref{a7}) between $l_0$ and $d_0$, since it is geometric. 
We will see that for a fast moving skater the contact length $l$ is 
substantially shorter than the $2l_0$ needed at rest. While for static
contact the total length, forward and backward, $2 l_0$ counts, for the
dynamic contact only the forward section $0 \leq x \leq l$ is relevant. 
What happens in the backward section $ -l \leq x < 0$ does not contribute
to the heat balance nor to the friction, since the contact between ice and
skate is broken.  

\section{The heat balance} \label{heat}

The heat generated by friction in the water layer leads to melting of 
ice. The first point for establishing the heat balance is to compute the
melting velocity $v_{\rm m} (x)$.  
The trough made by the skate has a width $w$ and a depth $d(x) + h(x)$.
So the trough grows downwards at a rate
\begin{equation} \label{b0}
v_{\rm tr} = \left(\frac{\partial [d(x,t) + h(x,t )]}
{\partial t} \right)_{t=0}.
\end{equation} 
Since the trough grows by melting with a velocity $v_{\rm m}$ and ploughing,
which has a downward velocity $v_{\rm ice}$, we have the equality
\begin{equation} \label{b1}
v_{\rm m} (x)+v_{\rm ice} (x)= v_{\rm tr}.
\end{equation} 
Working out the right hand side of Eq.~(\ref{b0}) gives the expression 
for the melting velocity
\begin{equation} \label{b2}
v_{\rm m} (x) = v_{\rm sk} (x) - v_{\rm ice} (x) - V \frac{\partial h}{\partial x}, 
\end{equation}
with $v_{\rm sk}$ given by Eq.~(\ref{a4}) and $v_{\rm ice}$ by Eq.~(\ref{a8}).

The main source of heat is the friction in the water layer due 
to the gradient in $v_x$. The gradient of the transverse flow $v_y$ contributes
an order of magnitude less to the heat generation.
So the frictional heat generated in a time $dt$ and a volume  
$h(x) \, w \, dx$ equals
\begin{equation} \label{b3}
d H (x) = \eta \frac{V^2}{h^2(x)} h(x) \, w \, dx \, dt.
\end{equation}
The heat gives rise to melting of a volume $d V (x)$, but it is a delicate
question which fraction of the heat is effective. There are two competitors
for melting. Inside the water layer a fraction $\zeta_{\rm w}$ will flow towards 
the ice and the remainder will flow towards the skate.
In Appendix \ref{water} it is shown that the fraction $\zeta_{\rm w} \geq 1/2$, 
but usually equal to 1/2, when the difference between skate and ice
temperature is small. The second competitor is
the heat flow inside the ice, which is a subtle point, playing a role at
low-temperature skating. We discuss this effect in Section \ref{temper}.
We stick here to the fraction 1/2 and get for the molten volume
\begin{equation} \label{b4}
d {\cal{V}} (x) = \frac{d H (x)}{2\rho L_H},
\end{equation}
with $\rho L_H$ the latent heat per volume. 
Equating this molten volume with the increase in water due to  
$v_{\rm m} (x)$ leads to the balance equation
\begin{equation} \label{b5}
v_{\rm m} (x) w  \, dx \, dt = d {\cal{V}} (x) = k
\frac{V}{h(x)} w \, dx \, dt,
\end{equation} 
where $k$ is the important parameter introduced by 
Le Berre and Pomeau \cite{pomeau} 
\begin{equation} \label{b6}
k = \frac{\eta V}{2 \rho L_H}.
\end{equation}
$k$ is a (microscopic) small length. We find for skating conditions 
$k = 2.1*10^{-11} $m.\footnote{Actually $k$ is about a factor 
$10^3$ smaller than the value 1.8*$10^{-8}$m given by the authors of 
\cite{pomeau}, since they erroneously
take for the water density $\rho=1$, while $\rho=10^3$ in SI-units.}.

We now turn this equation into a differential equation for $h(x)$ by 
substituting Eq.~(\ref{b2}) into Eq.~(\ref{b5}).
Bringing the difference $v_{\rm sk}-v_{\rm ice}$ to the right hand side yields
\begin{equation} \label{b7}
-V\frac{\partial h}{\partial x}  = k\frac{V}{h(x)} - 
[v_{\rm sk}(x)-v_{\rm ice}(x)].
\end{equation} 
This equation becomes useful if we have an expression for the
receding velocity $v_{\rm ice} (x)$. For the ice to recede, the water layer 
must have a pressure $p$ exceeding the hardness $p_H$ of ice. The pressure
in the water layer will be lower than $p_H$ near the midpoint $x=0$, 
where the layer is close to the open air. We will show that near the tip
$x=l$ the pressure will exceed $p_H$. We call the fraction with
$p>p_H$ the ploughing regime and the fraction with $p<p_H$ the melting regime.

In the melting phase we have $v_{\rm ice} (x)=0$ and with $v_{\rm sk} (x)$ from
Eq.~(\ref{a4}), we get the layer equation
\begin{equation} \label{b8}
-\frac{d h(x)}{dx} = \frac{k}{h(x)} - \frac{x}{R},
\end{equation}
which is the equation derived by Le Berre and Pomeau \cite{pomeau}.

In the ploughing phase we need the expression Eq.~(\ref{a8}) 
for the receding speed $v_{\rm ice}$. 
The pressure in the water layer has to depend on $y$, since it drives
out the water sideways. This causes the receding velocity to depend on $y$
and that in turn makes the layer thickness $h$ also dependent on $y$. In order
to stick to the approximation where $h$ depends only on $x$, we replace
Eq.~(\ref{a8}) by its average over $y$
\begin{equation} \label{b9}
v_{\rm ice} (x) = \gamma \int^{w/2}_{-w/2} \frac{dy}{w}  
[p(x,y,d(x)+h(x)) - p_H].
\end{equation}
In Appendix \ref{ydep} it is outlined how the $y$ dependence in $v_{\rm ice}$
can be accounted for. 

Eq.~(\ref{b8}) is derived without information
about the hydrodynamics of the water layer, other than that the gradient
in $v_x$ is the main source of friction. In the ploughing regime,
where $v_{\rm ice} (x) \neq 0$ we have to resolve the pressure dependence
from the flow pattern.

\section{The melting regime} \label{melting}

In order to analyse the layer equation (\ref{b8}), we introduce 
two length scales as a combination of the microscopic  length $k$
and the macroscopic length $R$. The longitudinal length
$s_l$ and the depth length $s_d$ are defined as
\begin{equation} \label{c1}
s_l = (k R^2)^{1/3}, \quad \quad {\rm and} \quad \quad 
s_d = (k^2 R)^{1/3}.
\end{equation} 
For the skating velocity $V=8$m/s, we have as the scale for the contact length
$s_l = 2.16$ mm and as scale for the thickness $s_d = 0.21 \mu$m.
Both are rather small.\footnote{The water layer thickness $s_h$ 
would multiply with a 
factor 100 for $\rho=1$ and the length $s_l$ with a factor 10. These values
are comparable with the values found by Le Berre and Pomeau \cite{pomeau}.}

If we use $s_l$ as a scale for the longitudinal coordinate $x$ and 
$s_d$ for the thickness $h$
\begin{equation} \label{c2}
x = s_l \, \bar{x} \quad \quad {\rm and} \quad \quad 
h = s_d  \, \bar{h},  
\end{equation}
Eq.~(\ref{b8}) becomes dimensionless 
\begin{equation} \label{c3}
-\frac{d {\bar{h}} (\bar{x})}{ d \bar{x}}=\frac{1}{\bar{h} (\bar{x})}-\bar{x}.
\end{equation} 
The advantage of this scaled equation is that no external parameters occur
in the equation. The skating velocity $V$ and radius of curvature $R$ 
come in via the scales $s_l$ and $s_d$ through the parameter $k$.
\begin{figure}[h]
\begin{center}
  \hspace{-3mm}\includegraphics[width=0.7\columnwidth]{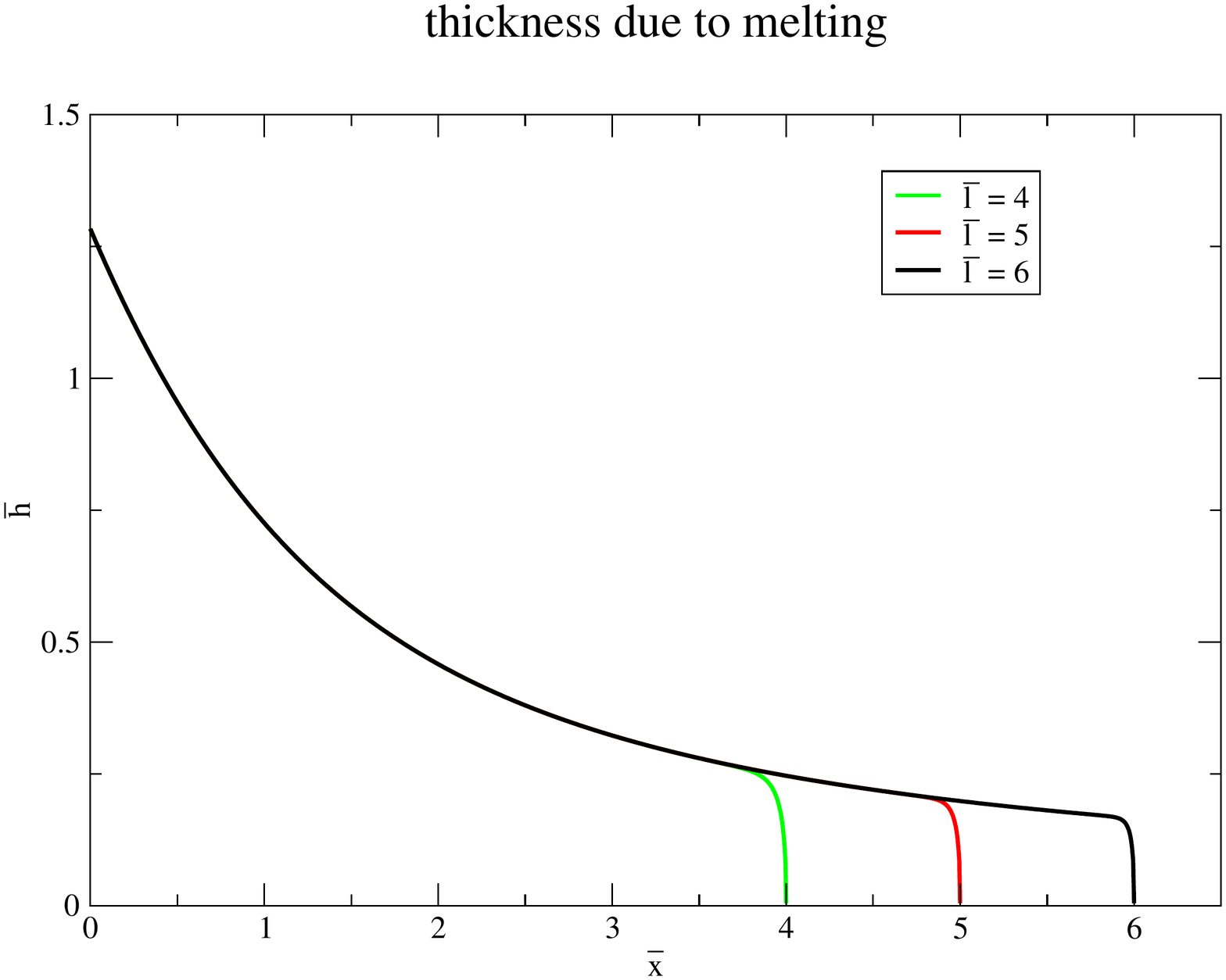}
  \caption{The scaled thickness $\bar{h}$ of the layer as a function of
the scaled position $\bar{x}$ in the melting regime. The curves are evaluated 
for some values of the scaled contact length $\bar{l}$. 
For negative $\bar{x}$ the water layer is irrelevant. It may be given
the constant value $\bar{h}(\bar{x})=\bar{h}(0)$.}
\label{melt}
\end{center}
\end{figure}

Eq.~(\ref{c3}) is easy to integrate numerically, starting from a guess for the
contact length $\bar{l}$. At $\bar{x}=\bar{l}$ the thickness $\bar{h}$ vanishes 
and thus the first term on the right hand side of Eq.~(\ref{c3}) dominates and
the solution behaves as
\begin{equation} \label{c4}
\bar{h} (\bar{x}) \simeq \sqrt{2 (\bar{l}-\bar{x})}, \quad \quad \quad
\bar{x} \rightarrow \bar{l}.
\end{equation} 
In Fig.~\ref{melt} we have given the curves for a few values of $\bar{l}$.
The curves distinguish themselves only near the tip $\bar{x}=\bar{l}$. 
Integrating the equation from below starting at $\bar{x}=0$, there is a value
$\bar{h}_0 \simeq 1.284$ such that the curves with $\bar{h}(0)>\bar{h}_0$
curve upwards asymptotically and the curves with $\bar{h}(0)<\bar{h}_0 $ 
bend downwards hitting the axis. 
The seperatrix starting at $\bar{h}(0)=\bar{h}_0$ behaves asymptotically as 
$\bar{h}(\bar{x}) \simeq 1/\bar{x}$.

The value of the contact length follows from the 
balance between the pressure in the water layer and
the weight $M$ of the skater, for which we need the pressure in the
water layer.

\section{The hydrodynamics of the water layer} \label{hydro}

The pressure is determined by the hydrodynamic equations of the water layer. 
The pressure distribution has been derived
both in \cite{lozowski} and \cite{pomeau}. Here we give the expressions
which are important for the next section. In Appendix \ref{velp} we sketch 
how the pressure follows from the assumption that the transverse flow has a 
Poisseuille form 
\begin{equation} \label{d1}
v_y (x,y,z) = a(x)\,  y \, [z-d(x)][h(x)-z+d(x)].
\end{equation}
The amplitude $a(x)$ determines, through the fluid equations,
the pressure behaviour. At the top and bottom of the layer we have
\begin{equation} \label{d2}
p(x,y,d(x)) = p(x,y,d(x)+h(x))= \eta a(x) \left( \frac{w^2}{4} - y^2 \right).
\end{equation}
The pressure is maximal in the middle of the skate blade and drops off 
towards the edges. The $y$ dependence of the pressure is
essential for pushing out the water towards the edges of the skate. (It
causes also an $y$ dependence in the layer thickness $h$, see Appendix
\ref{ydep}.) 

The incompressibility of water implies the connection of $a(x)$ with the
downward velocities of the top and bottom of the water layer 
\begin{equation} \label{d3}
v_{\rm sk} (x)-v_{\rm ice} (x) = a(x) h^3 (x)/6
\end{equation}
Eq.~(\ref{d3}) holds both in the melting and the ploughing phase.
In the melting regime, where $v_{\rm ice}=0$, it implies a simple relation
between $a(x)$ and $h(x)$
\begin{equation} \label{d4}
V \frac{x}{R} = a(x) h^3 (x)/6.
\end{equation} 
Using Eq.~(\ref{d4}) for $a(x)$ and Eq.~(\ref{d2}), gives for the 
average pressure in the 
melting phase the expression
\begin{equation} \label{d5}
\frac{1}{w} \int^{w/2}_{-w/2} dy p(x,y,d(x)+h(x)) = 
\frac{\eta w^2 V  }{R } \frac{x}{h^3 (x)}.
\end{equation} 
This presents a problem for the weight balance, if the melting phase
would apply all the way to the tip, where
$h(x)$ behaves as given by Eq.~(\ref{c4}). That 
leads to a diverging pressure, which is non-integrable. So some regularisation
near the tip is necessary, see \cite{pomeau}. In our treatment
this problem does not occur, since the the ploughing regime takes over 
as soon as the pressure exceeds the hardness $p_H$.

\section{The ploughing regime} \label{ploughing}

As follows from the analysis of the previous section, part of the
deformation of ice is due to the force on the ice. 
With Eq.~(\ref{d2}) and Eq.~(\ref{b9}) we find 
\begin{equation} \label{e1}
v_{\rm ice} (x) = \gamma [\eta a(x) w^2 /6 -p_H].
\end{equation} 
Using this expression in Eq.~(\ref{d3}) we
obtain the following relation between $a(x)$ and $h(x)$
\begin{equation} \label{e2}
a(x) = 6 \frac{V x/R +\gamma p_H}{h^3(x)+\gamma \eta w^2}. 
\end{equation} 
The heat balance equation (\ref{b7}) can be cast, with Eq.~(\ref{d3}), 
into the form
\begin{equation} \label{e3}
-\frac{d h(x)}{dx} = \frac{k}{h(x)} -\frac{a(x) \, h^3 (x)}{6V}.
\end{equation}
Then using $a(x)$ from Eq.~(\ref{e2}), turns it into an explicit 
layer equation for $h(x)$
\begin{equation} \label{e4}
-\frac{d h(x)}{dx} =  \frac{k}{h(x)} - 
\frac{x/R +\gamma p_H/V}{h^3(x)+\gamma \eta w^2} h^3(x).
\end{equation}
We note that putting $\gamma =0$, which is equivalent to putting $v_{\rm ice}=0$,
reduces indeed the equation to Eq.~(\ref{b8}) of the melting regime. On the
other hand, the limit $\gamma \rightarrow \infty$ reduces the equation to 
\begin{equation} \label{e5}
-\frac{d h(x)}{dx} = \frac{k}{h(x)} -
 \frac{p_H }{\eta w^2 V} h^3 (x),
\end{equation} 
which is the backbone of the equation derived by Lozowski and Szilder 
\cite{lozowski}. A very large $\gamma$ implies that the pressure at
the bottom of the layer stays equal to the hardness $p_H$ and that is
an implicit assumption in \cite{lozowski}. Eq.~(\ref{e5}) can be solved
analytically, see Appendix \ref{slow}.

In order to get a better insight in Eq.~(\ref{e4}), we make the equation 
dimensionless by introducing the same scaling as in Eq.~(\ref{c2}),
yielding the layer equation
\begin{equation} \label{e6}
-\frac{d \bar{h}}{d \bar{x}} = \frac{1}{\bar{h} (\bar{x})}- 
\frac{\bar{x} + c_1}{c_2+ \bar{h}^3 (\bar{x})} \bar{h}^3(\bar{x}),
\end{equation} 
with the dimensionless constants
\begin{equation} \label{e7}
c_1= \frac{\gamma p_H}{V} \left(\frac{R}{k}\right)^{1/3},\quad \quad \quad 
c_2=\frac{\gamma \eta w^2}{k^2 R}.
\end{equation} 
The magnitude of these constants depends on the value of $\gamma$, on which we
have little experimental evidence. With the value $\gamma p_H = 10^{-3}$
m/s, we get for skating conditions
\begin{equation} \label{e8}
c_1=1.27, \quad \quad \quad c_2=15.0 \quad \quad \quad 
c_3 = \frac{c_1}{c_2} =0.085.
\end{equation} 
Note that the ratio $c_3$ is independent of $\gamma$.

\section{The Cross-over from ploughing to melting}\label{cross} 

We must integrate Eq.~(\ref{e6}) starting from a value $\bar{l}$ till 
a point where the velocity $v_{\rm ice}(x)$ tends to become negative. 
Thus with Eq.~(\ref{e1}) we have to obey the condition 
\begin{equation} \label{f1}
\eta a(x) w^2 > 6 p_H.
\end{equation} 
With the expression (\ref{e2}) for $a(x)$ this translates to 
\begin{equation} \label{f2}
\eta w^2 V x/R > p_H\,  h^3 (x), \quad \quad {\rm or} \quad \quad
\bar{x} > c_3 \bar{h}^3 (\bar{x}).
\end{equation} 
At the top $\bar{x}=\bar{l}$ we have  $\bar{h} (\bar{l})=0$. So there the 
inequality is certainly fulfilled. At the midpoint $\bar{x}=0$, so 
there the inequality is certainly violated. Somewhere in between, at the
cross-over point $\bar{l}_c$, the ploughing regime merges smoothly 
into the melting regime. In dimensionless units, $\bar{l}_c$ is the 
solution of the equation
\begin{equation} \label{f3}
\bar{l}_c = c_3 \bar{h}^3 (\bar{l}_c).
\end{equation} 

At the cross-over point the layer thickness $\bar{h}_c=\bar{h} (\bar{l}_c)$
is the same in both regimes. The derivative is also continuous at the
cross-over point. We find in the ploughing regime
\begin{equation} \label{g5}
- \left(\frac{d \bar{h}}{d \bar{x}} \right)_{\bar{l}_c} = 
\frac{1}{\bar{h} (\bar{l_c})} - \frac{c_3 \bar{h}^3 (\bar{l}_c) + c_1}
{c_2 + \bar{h}^3 (\bar{l}_c)} \bar{h}^3 (\bar{l}_c) = 
\frac{1}{\bar{h} (\bar{l_c})} - c_3 \bar{h}^3 (\bar{l}_c) = 
\frac{1}{\bar{h} (\bar{l_c})} - \bar{l}_c,
\end{equation} 
which equals the value in the melting regime.

\section{Scaling the Pressure in the water layer} \label{pressure}

The pressure at the top of the water layer is given by Eq.~(\ref{d2}) and with 
Eq.~(\ref{e2}) for the amplitude $a(x)$ we get in the ploughing regime
\begin{equation} \label{g1}
p(x)= \eta w^2 \frac{V x/R +\gamma p_H}{\gamma \eta w^2+h^3(x)}.
\end{equation} 
It is interesting to compare this value with the hardness $p_H$ of ice and 
to express this ratio in dimensionless units
\begin{equation} \label{g2}
\bar{p} (\bar{x}) = \frac{p(x)}{p_H} = \frac{\eta w^2}{p_H/V} \, \frac{x/R 
+\gamma p_H/V} {\gamma \eta w^2+h^3(x)} = \frac{\bar{x} +c_1}
{c_1+ c_3 \bar{h}^3 (\bar{x})}.
\end{equation} 
This expression holds in the ploughing regime. In the melting regime we 
have
\begin{equation} \label{g3}
\bar{p} (\bar{x}) = \frac{1}{c_3} \, \frac{\bar{x}}{\bar{h}^3 (\bar{x})}.
\end{equation} 
Note that, with Eq.~(\ref{f3}),  both expressions (\ref{g2}) and 
(\ref{g3}) yield $\bar{p} (\bar{l}_c) = 1$. 
$\bar{p} (\bar{x})$ is larger than 1 in the ploughing phase and smaller 
than 1 in the melting phase. The maximum pressure occurs at the tip, 
$\bar{x}=\bar{l}$, where $\bar{h}=0$, with the value
\begin{equation} \label{g4}
\bar{p}_t = \bar{p} (\bar{l}) = 1 + \frac{\bar{l}}{c_1}.
\end{equation} 
As $\bar{l}$ will turn out to be around 6, this is a substantial ratio.

\section{The Macroscopic Forces} \label{forces}

The skate feels a normal and tangential force. The normal force
$F_N=Mg$ is the weight of the skater. The tangential friction force has 
two ingredients: the friction force $F_{\rm fr}$, due to the water layer and the 
ploughing force $F_{\rm pl}$, which pushes down the ice. All three forces
are related to integrals over the contact zone.  
The weight $M$ of the skater is balanced by the pressure at
the top in the water layer
\begin{equation} \label{j1}
F_N = w \int^l_0 dx p(x).
\end{equation} 
The friction force is given by the gradient of the flow in the water layer
\begin{equation} \label{j2}
F_{\rm fr} = \eta w \int^l_0 dx \frac{V}{h(x)} .
\end{equation}
The ploughing force results from the force that the pressure in the 
water layer exerts on the  ice in the forward direction. It is given by
\begin{equation} \label{j3}
F_{\rm pl} = w \int^l_0 dx \, \frac{x}{R} \, p(x).
\end{equation}
The ratio $x/R$ gives the component of the force in the forward direction.

Applying the scaling Eq.~(\ref{c2}) on $x$ and $h(x)$ and scaling the 
pressure with the hardness $p_H$, we get the expressions
\begin{equation} \label{j4}
\left\{ \begin{array}{rcl}
F_N        & = & a_N \displaystyle\int^{\bar{l}}_0 d \bar{x} \, \bar{p} 
( \bar{x}), \\*[4mm]
F_{\rm pl} & = & a_{\rm pl} \displaystyle\int^{\bar{l}}_0 d \bar{x} \, \bar{x} \, 
\bar{p} ( \bar{x}), \\*[4mm]
F_{\rm fr}   & = & \displaystyle a_{\rm fr} \int^{\bar{l}}_0 d \bar{x} \, 
\frac{1}{\bar{h}(\bar{x})}.
\end{array} \right.
\end{equation} 
The integrals are dimensionless and the constants have the dimension of a force
\begin{equation} \label{j5}
a_N = p_H\,  w\,  s_l, \quad \quad \quad a_{\rm pl} = p_H\,  w \, s_d, 
\quad \quad \quad a_{\rm fr}  = \eta \, V \, w  \frac{s_l}{s_d}.
\end{equation} 
Note that the ratio $a_{\rm pl}/a_N$ involves the ratio of the scales 
$s_d/s_l$, which is a reflection of the fact that the normal force acts 
over the longitudinal length $l$ and the ploughing over the depth $d$. 
In order to compare friction with ploughing, we use the number
$\lambda$ introduced in Eq.~(\ref{B5}), leading to 
\begin{equation} \label{j6}
\eta V = 2 k \,\rho L_H = 2 k \, \lambda \, p_H.
\end{equation} 
This gives for the relation between $a_{\rm fr}$ and  $a_{\rm pl}$
\begin{equation} \label{j7}
a_{\rm fr}= 2 k \,\lambda \, p_H \, w \, \frac{s_l}{s_d}= 
2 p_H  w \, \lambda s_d  = 2 \lambda \, a_{\rm pl}.
\end{equation} 

An interesting feature of pressures $p(x)$, exceeding the hardness $p_H$
in the ploughing regime, is that they shrink the contact length $l$ and
the penetration depth $d$, since $d$ goes with the square of $l$. 
So the skater ``rises'' due to his velocity. We find in the limit 
$V \rightarrow 0$ an indentation depth $d \simeq 44 \mu$m and for $V=8$m/s
a value $d=4.5 \mu$m.\footnote{See Appendix \ref{slow} for the relation 
between the static $d_0$ and $d$ in the slow limit.} 
For slow velocities  the $F_{\rm fr}$  vanishes and $F_{\rm pl}$ has a 
limit $\simeq 0.7$ N for a skater of 72 kg. For the $V=8$m/s we find
$F_{\rm fr} =0.84$ N and $F_{\rm pl} = 0.29$ N. 
So the large pressure build-up near the tip, reduces the ploughing force,
from dominant at $V \rightarrow 0$, to a fraction of the total friction force.

\section{Velocity dependence of the friction}\label{velocity}
  
\begin{figure}[h]
\begin{center}
  \hspace{-3mm}\includegraphics[width=0.7\columnwidth]{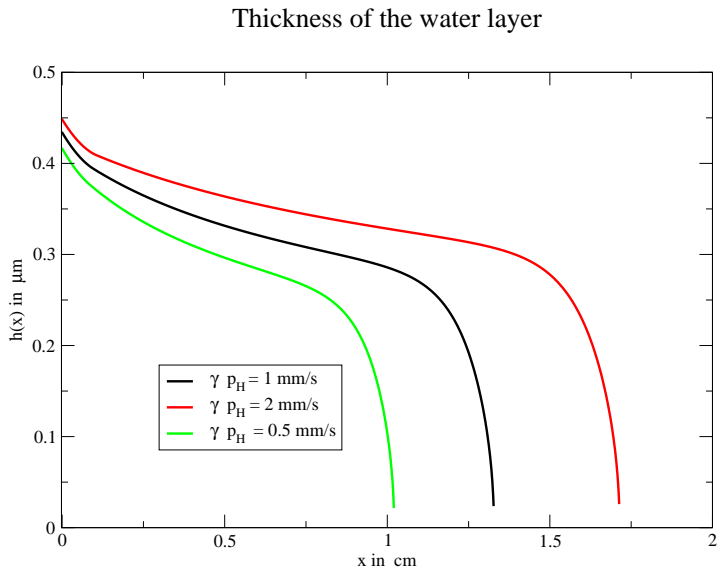}
  \caption{The shape of the water layer for some values of $\gamma$ and skating 
conditions}
\label{layer}
\end{center}
\end{figure}

The integration of the layer equation is straightforward once we know the
contact length $l$. The value of $l$ determines the weight 
of the skater. Since the 
weight is given, we must find the contact length by trial and error. 
In Fig.~\ref{layer} we have drawn the shape of the water layer for a few
values of the deformation rate $\gamma$ and a skater weight of 72 kg.
The curves end at $x=l$ and one observes that the contact length is
rather sensitively dependent on the value of $\gamma$. This is not 
surprising since $\gamma$ has a direct influence on the pressure in
the water layer and the pressure determines the weight. The small 
up-swing of the thickness in the middle of the skate ($x=0$) is a 
manifestation of the melting phase. On the other hand the overal thickness
of the layer does not depend strongly on the value of $\gamma$.

\begin{figure}[h]
\begin{center}
  \hspace{-3mm}\includegraphics[width=0.7\columnwidth]{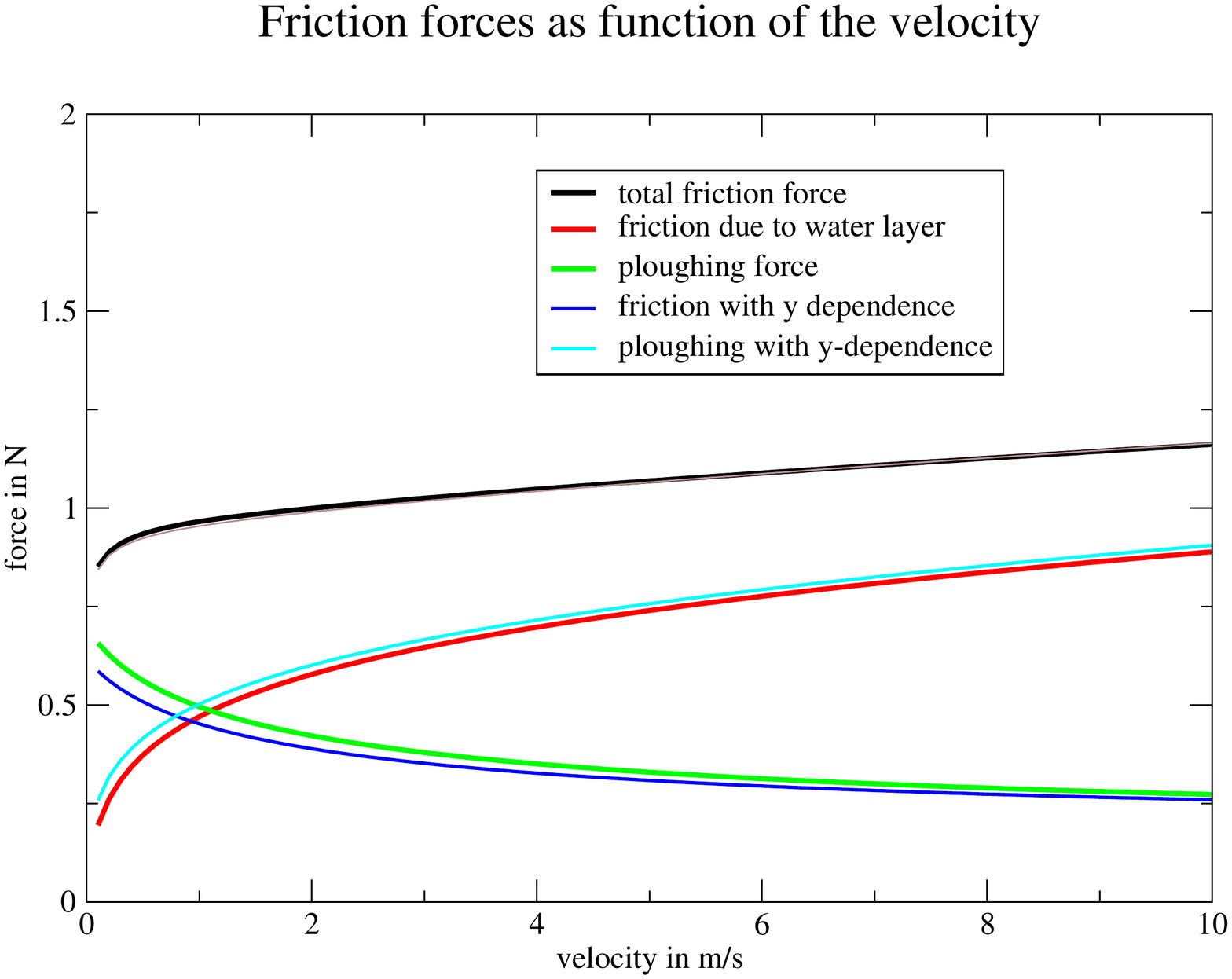}
  \caption{The various contribution to the friction as function
of the velocity $V$, for otherwise skating conditions.}
\label{nfriction}
\end{center}
\end{figure}

The next result is the friction as function of the velocity. 
In Fig.~\ref{nfriction} we have drawn how the ploughing and water friction
combine to the total strength of the friction. While both components
vary substantially with the velocity, the combination is remarkably
constant over a wide range of velocities. One
observes that the low $V$ limit (exhibiting a square root dependence on $V$),
covers only a very small region of velocities. 
In the Fig.~\ref{nfriction} we have also plotted the influence on the 
contributions, if one
takes the $y$ dependence of the thickness into account. The effects on 
friction and ploughing are small and opposite, such that 
the change of the total friction is not visible in the Fig.~\ref{nfriction}.

\begin{figure}[h]
\begin{center}
  \hspace{-3mm}\includegraphics[width=0.7\columnwidth]{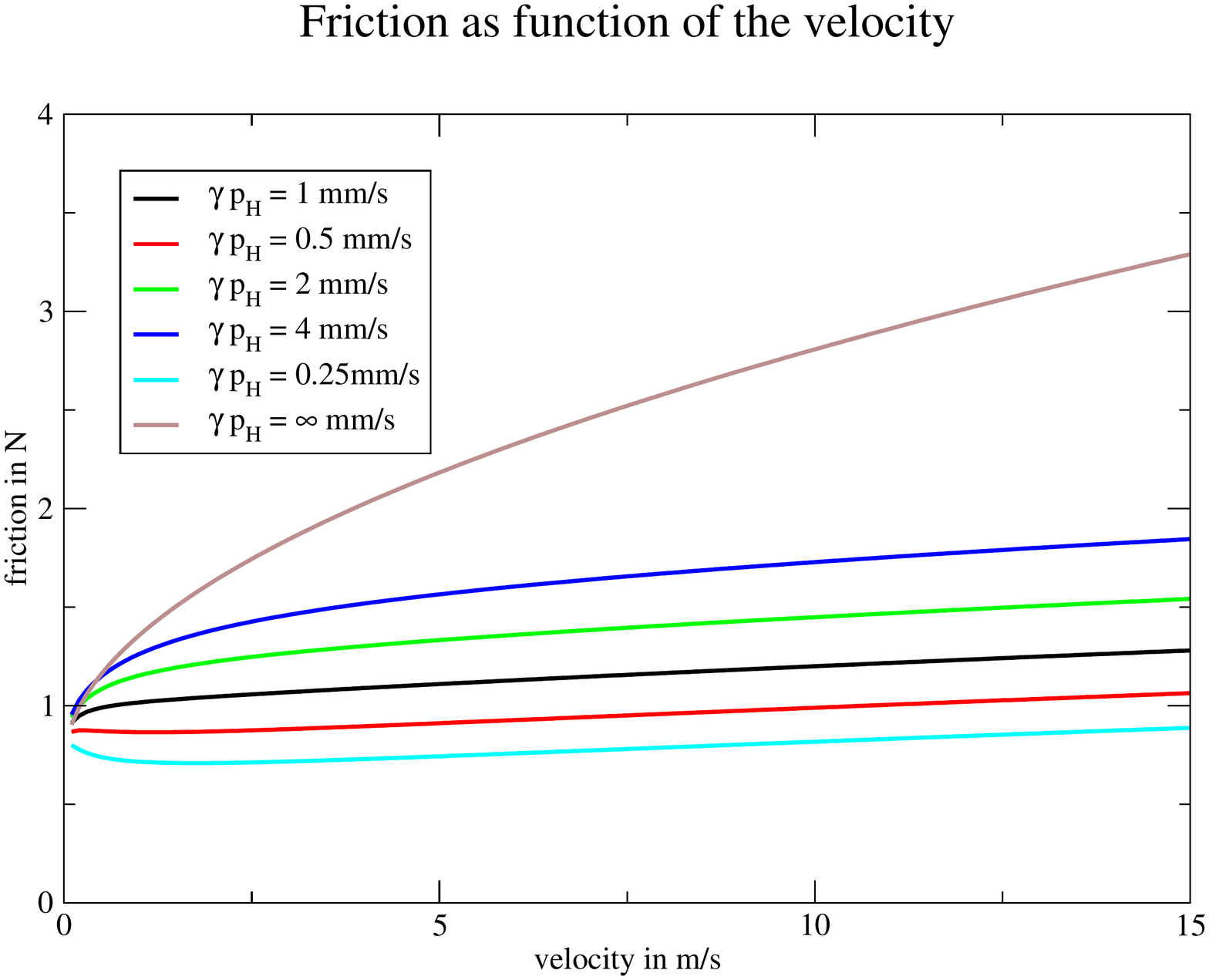}
  \caption{The total friction as function of the velocity
 for some values of $\gamma$}
\label{friction}
\end{center}
\end{figure}

In order to see how much the value of $\gamma$ influences the friction, we
have drawn in Fig.~\ref{friction} the total friction as function of the
velocity for some values of $\gamma$. The influence of $\gamma$ is noticeable,
but not dramatic. A factor 16 difference in $\gamma p_H$, between 
$\gamma p_H=4$ mm/s and $\gamma =0.25$ mm/s,
gives a factor 2 in the friction for large velocities.
But there is a substantial difference with respect to the 
theory of Lozowski and Szilder \cite{lozowski}, using $\gamma=\infty$.

Usually the friction is expressed in terms of the friction coefficient 
$\sigma$, being the ratio 
of the tangential and the normal force. In the present case it reads
\begin{equation} \label{h1}
\sigma = \frac{F_{\rm fr} + F_{\rm pl}}{F_N} .
\end{equation}
However, for skating the friction coefficient is not independent of the 
normal force. In a standard friction experiment the contact surface is 
proportional to the normal force and the friction force is proportional to 
the contact area, such that in the friction coefficient the contact area 
drops out. This proportionality does not hold for skating. 
The order of magnitude of the friction coefficients is 0.002 for skating
conditions. We estimate the contact area for skating conditions as
$s_l \, \bar{l} \,  w \simeq 14.3$ mm$^2$.

\section{Temperature dependence of the friction} \label{temper}

So far we have considered temperatures close to the melting point of ice,
where temperature gradients and associated heat flows are small. At lower
temperatures they start to play a role. 
In order to melt ice, one first has to heat it to the melting temperature 
$T_m$. If the difference between the melting temperature and the surface
temperature $T_s$ is positive, i.e.~when the surface
temperature is lower than the melting temperature, one has to increase
the latent heat $L_H$ with the amount needed to heat the ice. 
Since the latent heat is 80 times the heat 
necessary to raise the temperature of ice by one degree, 
this is usually a small correction. 

Another small
correction comes from the heat flux which may exist in the ice layer. In a
skating rink the ice is cooled from below and there is a heat flow downwards.
Natural ice freezes by cooling the top layer and correspondingly the heat
flow is upwards. But the temperature gradients are small with respect to
the temperature gradients in the water layer, so the effect on the
amount of ice that melts is small and we leave it out.

However, as pointed out in \cite{lozowski}, there is another heat flow, 
which can have an important effect on the friction 
at low ice temperatures. If the surface temperature is low, one has to heat 
the surface, before it melts. This causes a temperature gradient in the
ice and an associated heat leak into the ice. The melting occurs under
pressure and one has to raise the temperature, not to zero centigrade, but
to the melting temperature $T_m$ at that pressure. 
Since the pressure $p$ in the water layer is large, $T_m$
can be substantial below zero degree centigrade. 
The lowering of the melting temperature is approximately given by
\begin{equation} \label{k1}
T_m = - 0.1 p*10^{-6}.
\end{equation}
The maximum value of $T_m$ occurs at $p=2*10^8$ Pa, producing a $T_m$
of -20 degrees centigrade. Since the pressure varies strongly with the
position $x$ of the contact, $T_m$ varies also with $x$.
In the middle of the skate, where the contact ends, the pressure vanishes
and the melting temperature $T_m(0)=0$. At the tip the pressure is maximal 
and $T_m(l)$ reaches its lowest point. We have to distinguish two cases:
$T_m(l)<T_s$ and $T_m(l)>T_s$. In the former case, there is a point $x_0$ where
$T_m(x_0)=T_s$. For $x>x_0$ the melting temperature is then below the
surface temperature and no heat is needed to raise the ice to $T_m$. In the
latter case the ice is heated all along the contact line and at the tip a 
sudden jump in the surface temperature occurs. 

In Appendix \ref{heatice}
we have given the derivation of the temperature gradient in the ice
at the surface. It reads
\begin{equation} \label{k2}
\left(\frac{ \partial T(x,z)}{\partial z}\right)_{z=0} = 
\left(\frac{V} {\pi \alpha_{\rm ice}}\right)^{1/2} \left(
-\int_x^{x_0} dx' \frac{1}{\sqrt{x'-x}} \frac{\partial T_m(x') }{\partial x'}+
\frac{T_m(l)-T_s}{\sqrt{l-x}} \right)
\end{equation} 
The understanding is that the last term is absent for $T_m(l)<T_s$ and in
the other case the integral extends to $x_0=l$. 
In \cite{lozowski} only this last term is taken into account, together with
setting  $T_m(l)=0$.

The gradient gives a downward heat flow at the surface $z=0$
\begin{equation} \label{k3}
J_{\rm ice} (x) = -\kappa_{\rm ice} \left(\frac{\partial T(x,z)}
{\partial z}\right)_{z=0}.
\end{equation} 
This gradient takes away a fraction of the heat supplied by $J_{\rm w-ice}$ 
\begin{equation} \label{k4}
\zeta(x)  = 1 - \frac{J_{\rm ice}(x)}{J_{\rm w-ice}},
\end{equation} 
with $J_{\rm w-ice}$ given by Eq.~(\ref{D10}).

\begin{figure}[h]
\begin{center}
  \hspace{-3mm}\includegraphics[width=0.7\columnwidth]{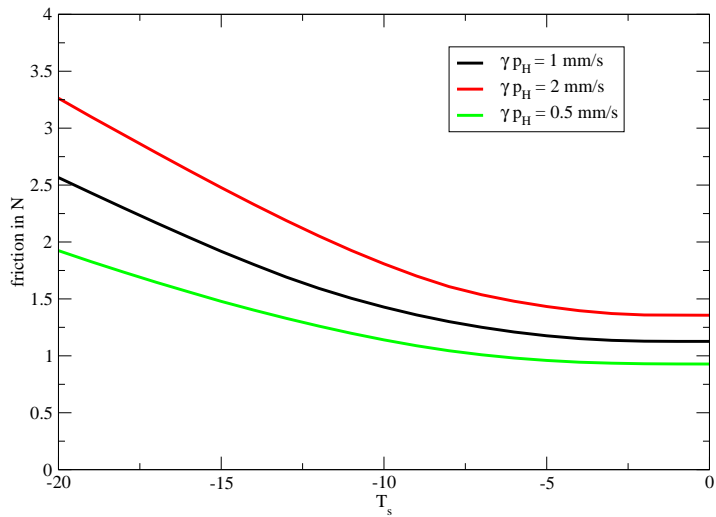}
  \caption{The friction for skating conditions as function of the temperature.}
\label{stemper}
\end{center}
\end{figure}

In the layer equation we have to replace the first term of the right hand side
by
\begin{equation} \label{k5}
\frac{k}{h(x)} \rightarrow \frac{k \zeta(x)}{h(x)}.
\end{equation}
In Fig.~\ref{stemper} we have drawn the friction as function of the 
surface temperature for skating conditions. Note that the friction hardly
changes in the region $0>T_s>-5^0$C, after which the friction starts to 
increase. The influence of $\gamma$ is similar for all temperatures. 

\section{Discussion}\label{discussion}

We have investigated the thickness of the water layer underneath the
skate as a result of melting of ice by the frictional heat. 
In skating two processes take place: a plastic 
deformation of the ice (ploughing) and the generation of a water layer
by melting. For low velocities ploughing dominates  and for
high velocities friction in the water layer dominates. 
In the skating range of velocities, 
the total friction is rather independent on the velocity. 
The friction in the water layer increases
with the velocity, which is compensated by an almost equal decrease in the
ploughing force. A high skating velocity causes a high pressure in the 
water layer, lifting the skater. Consequently the skate penetrates less 
deep into the ice and the ploughing force decreases. The theory assumes
that the thickness of the water layer is large enough to treat it 
hydrodynamic system. For low velocities and low temperatures this
assumption breaks down (see Appendix \ref{heatice}). 

There are two important material constants of ice, 
determining the friction: the hardness $p_H$ and
the deformation rate $\gamma$. Unfortunately no accurate data exist on
these constants, which hampers a quantitative calculation of the friction.
In particular the value of $\gamma$ is poorly known, while it has a substantial
influence on the magnitude of the friction. The relevance of $\gamma$ becomes
clear from an estimate of the speed at which ice has to be pushed down at
the tip of contact. For a forward velocity of 10 m/s, the downward speed of
the ice is about 1 cm/s. Such high deformation rates require large pressures, 
several times the hardness.

The most important theoretical parameter is $k$, defined in Eq.~(\ref{b6}),
which is microscopically small for reasonable values of the velocity of the 
skate. In combination with the macroscopic curvature $R$ of the skate, 
two length scales follow: the longitudinal scale $s_l$  and the depth 
scale $s_d$ defined in Eq.~(\ref{c1}). $s_l$ is a measure for the contact 
length and $s_d$ gives the magnitude of the thickness of the 
water layer. 

Our analysis
combines elements of the theory of Le Berre and Pomeau \cite{pomeau}, which
only accounts for the effects of melting and the theory of Lozowski and 
Szilder, which equals the pressure in the water layer  to the hardness
of ice. The new element is that we propose that the ice recedes with a velocity 
proportional to the excess pressure with respect to the hardness. In the 
theory of Le Berre and Pomeau the ice does not recede (which is equivalent with
$\gamma=0$), in spite of the fact
that, in their approach, the pressure grows unlimitedly near the tip. 
In the theory of Lozowski and Szilder the ice adapts instantaneously (which is
equivalent with $\gamma=\infty$),
keeping the pressure equal to the hardness.

We have mainly considered skating near the melting point. At low temperatures
a number of new elements come into play, which we have indicated in Section 
\ref{temper}. A quantitative discussion of these effects is delicate, since
they depend not only on the conditions of the ice, but also on the value of
the constant $\gamma$ in the Bingham Eq.~(\ref{a8}). Since the
hardness $p_H$ and the deformation rate $\gamma$,
are not very well known as function of the temperature, a precise measurement 
of these properties would be very welcome. 

We have left out a number of refinements in order to focus on the essential
features of skating. Refinements that can be treated in the presented context
are:
\begin{itemize}
\item We have omitted the influence of the melting of the ice at the
sides of the skate. A simple treatment adds to the width $w$ on both sides
the amount $d(x)$. Since the indentation depth $d$ of the skate
is very small compared to the width $w$ of the blade (we find a ratio 1/500)
it gives a small correction.
\item We have assumed that only the gradient of the forward velocity 
contributes to the friction and the corresponding heat generation. It is
easy to take into account the contributions of the gradient
in the transverse velocity. The relative importance of the longitudinal
and transverse heat generation is of the order $1/\lambda$, see Eq.~(\ref{B5}). 
This means that the transverse velocity gradient contributes only a few 
percent to the generated heat.
\item Most of our calculations are based on the assumption that the 
thickness $h$ depends only on the longitudinal coordinate $x$. In Appendix
\ref{ydep} we have made a start of taking the transverse $y$ dependence
into account. A fully consistent treatment, including the hydrodynamic
equations, is computationally quite involved and as far as the friction
is concerned not very encouraging, as the effect is quite small 
in lowest order (see Fig. \ref{nfriction}).
The reason is that the variation of 
$h(x,y)$ with $y$ is modest except at the edges of the skate. 
\end{itemize}

There are several influences outside our scope, 
such as the humidity of the air and the 
addition of suitable chemicals to the surface layer, which are important 
for speed skating records, but not essential for the phenomenon of skating.
Apart from a more accurate measurement of $p_H$ and $\gamma$,
it would be interesting if the deformation of the ice, due to the skate,
could be observed. Presumably the 10\% difference in density between
ice and water, which we ignored, plays an important role for the form of
the deformation.

De Koning et al.~\cite{schenau} report a friction force of 3.8 N for the
straight strokes and 4.9 N for the curves. The difference is due to the fact 
that in the curves the skate is at an angle with the ice. In the straights
there are also parts, at the begin and end of the stroke, where the skate
makes an angle with the ice. So for the upright part, for which we perform
the calculation, one estimates a friction force around 2 N. This compares
well with the values we see in Fig. \ref{friction}. A fit might be seen as 
a measurement of $\gamma$ and tends to the value $\gamma p_H = 2$ mm/s.

In Appendix \ref{slow} we discuss the slow velocity limit $V \rightarrow 0$, 
which is hardly relevant for skating, but may be useful for measurements 
in the laboratory, involving low $V$.

{\bf Acknowledgments.} The author is indebted to Tjerk Oosterkamp for 
drawing his attention to the problem and for careful reading of the manuscript
and to Tom van de Reep for explaining
the details of the measurements in Leiden. He also acknowledges discussions
with the experimental group of Daniel Bonn in Amsterdam, in particular the
discussions with Bart Weber on ploughing. Numerous conversations about the
properties of ice with Henk Bl\"ote are highly appreciated.
\appendix

\section{Velocity and Pressure in the water Layer} \label{velp}

In this Appendix we discuss the hydrodynamics in the water 
layer underneath the skate. We take advantage of the fact that we have
three different length scales: in the $x$ direction the scale is in 
centimeters, in the $y$ direction in millimeters and in the $z$ direction 
in microns. So the gradients in the $z$ direction are much larger than in
the other directions and we may use the lubrication approximation of 
the Navier-Stokes equations for an incompressible fluid
\begin{equation} \label{A1}
\nabla p = \eta \, \Delta {\bf v}, \quad \quad {\rm and} \quad \quad 
\nabla \cdot {\bf v} = 0.
\end{equation}  
The velocity in the $x$-direction is forced by the motion of the skate
\begin{equation} \label{A2}
v_x = V \left(1 - \frac{z-d(x)}{h(x)} \right).
\end{equation} 
At the top of the layer $z=d(x)$, the velocity of water equals that of the 
skate and at the bottom, $z=d(x)+h(x)$, it vanishes at the solid ice surface.

The velocity in the $y$ direction has a Poisseuille form
\begin{equation} \label{A3}
v_y (x,y,z) = a(x)\,  y \, [z-d(x)][h(x)-z+d(x)],
\end{equation} 
This velocity component vanishes at the skate blade $z=d(x)$ as well as
at the bottom of the layer at $z=d(x)+h(x)$. The linear dependence on $y$
is a consequence of the incompressibility of water. To see this, consider a 
volume between $x$ and $x+\delta x$, $y$ and $y+\delta y$ and $z=d(x)$ and 
$ z=d(x)+h(x)$. At the top it goes down with 
the velocity $v_{\rm sk} (x)$ and at the bottom it may go down with a velocity 
$v_{\rm ice} (x)$.  The total decrease of the volume
due to vertical motion of the top and bottom boundary equals
\begin{equation} \label{A4}
\Delta V_v = [v_{\rm sk} (x)-v_{\rm ice} (x)] \delta x \,\delta y\,\delta t.
\end{equation}  
In the horizontal direction we have an inflow at $y$ and an outflow at 
$y+\delta y$ resulting in the net displaced volume
\begin{equation} \label{A5}
\int^{h(x)+d(x)}_{d(x)} dz [v_y (x,y+\delta y,z) -  v_y (x,y,z)] 
\delta x \delta t =  a(x) \, h^3(x) \, \delta y  \delta x \delta t /6.
\end{equation} 
As water is incompressible we have the balance
\begin{equation} \label{A6}
v_{\rm sk} (x)-v_{\rm ice} (x) = a(x) h^3 (x)/6.
\end{equation} 
The linear dependence of $v_y$ on $y$ makes the right hand side 
in Eq.~(\ref{A5}) independent of $y$.

The third component of the velocity is given by
\begin{equation} \label{A7}
v_z =v_{\rm ice}(x)+ a(x) \left(\frac{[z - d(x)]^3}{3} - 
\frac{[z-d(x)]^2 h(x)}{2} + \frac{h^3 (x)}{6} \right).
\end{equation} 
Note that we have chosen the constants such that at the top  
$v_z(x,y,d(x))=v_{\rm sk}(x)$ and at the bottom 
$v_z(x,y,d(x)+h(x))=v_{\rm ice}(x)$.

The pressure distribution compatible with this flow field is
fixed up to a constant. Here we take the boundary condition 
\begin{equation} \label{A7}
p(x,w/2,d(x)) = 0,
\end{equation} 
using that at the corners of the furrow the pressure is (nearly) 
zero. This gives the pressure the form
\begin{equation} \label{A8}
p(x,y,z) = \eta \, a(x) \left( \frac{w^2}{4} - y^2 - [z - d(x)] 
[d(x) +h(x)-z] \right).
\end{equation} 
At the top and the bottom the pressure equals
\begin{equation} \label{A9}
p(x,y,d(x)) = p(x,y,d(x)+h(x))= \eta a(x) \left( \frac{w^2}{4} - y^2 \right).
\end{equation} 
which is maximal in the middle of the skate blade.

It is easy to verify that the flow field and the pressure fulfil the 
Navier-Stokes equations (\ref{A1}),
provided that we consider for the differentiation only the explicit 
$y$ and $z$ dependence and ignore the $x$ dependencies of
$a(x)$ and $h(x)$ for the calculation of the gradients.

\section{The $y$ dependence of the water layer} \label{ydep}

We see from Eq.~(\ref{A8}) that the pressure depends explicitly 
on $y$. This implies, through the expression (\ref{a8}) for 
$v_{\rm ice}$, that also $v_{\rm ice}$ is dependent on $x$ and $y$. That in
turn forces the function $a$ and $h$ to depend also on $x$ and $y$.
Taking the $y$ dependence fully into account, also for the detailed solution 
of the hydrodynamic equations in the layer of varying thickness, is quite 
involved. Here we give a first step, which focusses on the explicit $y$
dependence that enters into the equations.
With Eq.~(\ref{a8}) and (\ref{A8}) one has
\begin{equation} \label{i1} 
v_{\rm ice} (x,y) = \gamma[\eta a(x,y)(w^2/4-y^2) - p_H].
\end{equation} 
If $v_{\rm ice}$ depends on $x$ and $y$, we also must change the layer 
Eq.~(\ref{e5}) into
\begin{equation} \label{i2}
-\frac{\partial h(x,y)}{\partial x} = \frac{k}{h(x,y)} - 
\frac{1}{V} [v_{\rm sk} (x) - v_{\rm ice} (x,y)].
\end{equation}
Finally the connection between $a$ and $v_{\rm ice}$, as given by 
Eq.~(\ref{A5}), changes into
\begin{equation} \label{i3}
v_{\rm sk} (x) - v_{\rm ice} (x,y) = \int^{h(x,y)+d(x)}_{d(x)} dz 
\frac{\partial v_y}{\partial y} = \frac{1}{6} \, \frac{\partial}{\partial y} 
 a(x,y) y h^3 (x,y).
\end{equation}
The last step uses that $v_y$ vanishes at the boundaries.
The three equations (\ref{i1})-(\ref{i3}) determine the behaviour of the
three quantities $a(x,y), h(x,y)$ and $v_{\rm ice} (x,y)$. We first eliminate
$v_{\rm ice}$ by inserting Eq.~(\ref{i1}) into Eqs. (\ref{i2}) and 
(\ref{i3}), which leads to the set
\begin{equation} \label{i4}
\left\{ \begin{array}{rcl}
\displaystyle -\frac{\partial h(x,y)}{\partial x} & = & \displaystyle 
\frac{k}{h(x,y)} - \left(\frac{x}{R} + \frac{\gamma p_H}{V}\right) + 
\frac{\gamma \eta}{V} a(x,y) [w^2/4-y^2], \\*[4mm]
\displaystyle \frac{x}{R} + \frac{\gamma p_H}{V} & = & \displaystyle
\frac{1}{6V} \, \frac{\partial}{\partial y}  a(x,y) \, y \, h^3 (x,y)+ 
\frac{\gamma \eta}{V} a(x,y) [w^2/4-y^2].
\end{array} \right.
\end{equation}

The equations simplify in the center $y=0$ where we may use
\begin{equation} \label{i5}
\frac{\partial}{\partial y}  a(x,y) \, y \, h^3 (x,y) \simeq 
 a(x,y) h^3 (x,y).
\end{equation}
In the anticipation that the variation of $a$ and $h$ with $y$ is modest,
we use the approximation (\ref{i5}) for the whole width. Then $a(x,y)$ can
be expressed in terms of $h(x,y)$ as
\begin{equation} \label{i6}
a(x,y)=\frac{6(V x/R +\gamma p_H)}{h^3(x,y)+\gamma \eta [3w^2/2-6y^2]}
\end{equation} 
 
\begin{figure}[h]
\begin{center}
  \hspace{-19mm}\includegraphics[width=0.7\columnwidth]{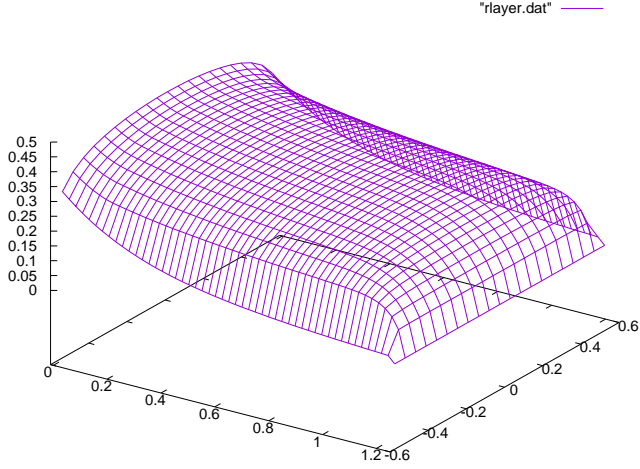}
  \caption{The vertical axis gives the thickness $h(x,y)$ of the water layer
in $\mu$m. In the horizontal direction the (longer) $x$ coordinate
is measured in cm and the $y$ coordinated in mm.} 
\label{rlayer}
\end{center}
\end{figure}

For calculational purpose we give the scaled version of the equations, 
using for $y$ and $a$ the scaling
\begin{equation} \label{i7}
y=w \, \bar{y}, \quad \quad \quad a(x,y) = \frac{V}{k^2 R}
 \left(\frac{k}{R} \right)^{1/3} \bar{a}(\bar{x},\bar{y}).
\end{equation}
For $\bar{a}$ expression (\ref{i6}) becomes 
\begin{equation} \label{i8}
\bar{a}(\bar{x},\bar{y}) = \frac{6(\bar{x}+c_1)}{\bar{h}^3(\bar{x},\bar{y})+
c_2(3/2 -6 \bar{y}^2)}.
\end{equation}  
The constants $c_1$ and $c_2$ are defined in Eq.~(\ref{e7}). With this value
of $a$ inserted into the first Eq.~(\ref{i4}), we get for $\bar{h}$
the equation 
\begin{equation} \label{i9}
-\frac{\partial \bar{h}(\bar{x},\bar{y})}{\partial \bar{x}}  = 
\frac{1}{\bar{h}(\bar{x},\bar{y})} - \frac{(\bar{x}+c_1)
\bar{h}^3(\bar{x},\bar{y})} {\bar{h}^3(\bar{x},\bar{y})+c_2(3/2 -6 \bar{y}^2)}.
\end{equation} 
This equation has to be solved, starting from a value $\bar{x}=\bar{l}$,
where the thickness behaves as indicated in Eq.~(\ref{c4}).
The transition to the melting regime occurs at
\begin{equation} \label{i10}
\bar{x} (3/2- 6 \bar{y}^2) = c_3 \bar{h}^3(\bar{x},\bar{y}).
\end{equation} 
From there on the equation reads in the melting regime as before
\begin{equation} \label{i11}
-\frac{\partial \bar{h}(\bar{x},\bar{y})}{\partial \bar{x}}  = 
\frac{1}{\bar{h}(\bar{x},\bar{y})} - \bar{x}.
\end{equation} 
In contrast to the equation where the average pressure was employed, the
transition from the ploughing to the melting regime is $y$ dependent.
It occurs immediately at the edges $\bar{y}=\pm 1/2$ and lastly in the
middle $\bar{y}=0$.
In Fig.~\ref{rlayer} we have plotted the solution of Eqs.~(\ref{i9}) and
(\ref{i11}) for skating conditions. 
Only at the edges there is a substantial $\bar{y}$ 
dependence. It is a consequence of the boundary condition that the pressure
should vanish at the edges of the skate. In fact the pressure is always 
higher than the atmospheric pressure, 
but that is a small value as compared to the hardness
of ice, which is the scale for the pressure in the water layer.

The approximation Eq.~(\ref{i5}) can be improved by computed the derivatives
of $\bar{a}$ and $\bar{h}$ from the solution of Eqs.~(\ref{i9}) and (\ref{i11})
and adding that as a correction to Eq.~(\ref{i5}). 
In view of the small influence on the
friction by the first approximation outlined in this section, 
(see Fig. \ref{friction}), such a further refinement is not worth while. 

\section{The slow velocity limit} \label{slow}

In this Section we discuss the limit of the velocity $V \rightarrow 0$.
When the velocity $V$ of the skate becomes small, the scaling used in the
previous sections is not adequate because the scales $s_l$ and $s_d$ 
vanish in the limit of $V \rightarrow 0$. The constants $c_1$ and $c_2$,
on the other hand start to diverge as
\begin{equation} \label{B1}
c_1 \sim V^{-4/3}, \quad \quad \quad  {\rm and} \quad \quad \quad c_2 \sim V^{-2}.
\end{equation}  
Using these limits in Eq.~(\ref{g2}) for $\bar{p}$ we see that $\bar{p}$
approaches 1, implying that for low velocities the pressure in the
water layer hardly rises above the hardness $p_H$. Consequently the ice will
recede also slowly. But if the pressure equals $p_H$, Eq.~(\ref{e5})
of Lozowski and Szilder \cite{lozowski} becomes valid.
 
Fortunately Eq.~(\ref{e5}) can be solved exactly.  Using that the water layer
vanishes at the top $x=l$ of the skate yields the expression for the layer
\begin{equation} \label{B2}
h (x) = A [\tanh ((l-x)/l_a]^{1/2}.
\end{equation} 
The asymptotic thickness $A$ of the layer is given by
\begin{equation} \label{B3}
A =  \left(\frac{\eta^2 w^2}{p_H \rho L_H}\right)^{1/4}  V^{1/2} = 
(k^2 \lambda)^{1/4}
\end{equation}
and the length $l_a$ of the onset of the asymptotic value reads
\begin{equation} \label{B4}
l_a = \frac{w}{2} \left(\frac{\rho L_H}{p_H}\right)^{1/2}= \frac{w}{2}
\lambda^{1/2}.
\end{equation}
The ratio 
\begin{equation} \label{B5}
\lambda= \rho L_H/p_H=22.72,
\end{equation} 
is a number, yielding $l_a=2.62$mm.

In principle, we still have to match this solution with the solution in
the melting regime. However, for $V \rightarrow 0$ the melting regime
shrinks to zero and the solution Eq.~(\ref{B2}) applies to the whole region.

In the low velocity limit the expressions simplify, since the
pressure in the layer approaches $p_H$. Going back to the first three
expression (\ref{j1})-(\ref{j3}), we have the equation for $l$
\begin{equation} \label{B6}
F_N = Mg = p_H w l, \quad \quad {\rm or} \quad \quad l= \frac{F_N}{p_H w}
\end{equation} 
and $l$ becomes equal to the static contact length $2 l_0$. The indentation
depth $d$ approaches therefore 4 $d_0$, with $d_0$ the static value.
The ploughing force reads in the limit $V \rightarrow 0$
\begin{equation} \label{B7}
F_{\rm pl} = p_H w d = \frac{p_H w l^2}{2 R} = \frac{F^2_N}{2 p_H w R},
\end{equation} 
using $l$  from Eq.~(\ref{B6}). This is an interesting relation. At
zero velocity there is no water layer and the ploughing force is the
only friction. It shows that Amonton's law does not
hold, since the friction is not proportional to the normal force. 
Note that the relation contains only the hardness $p_H$ and that it is
therefore a relation to measure the hardness.
 
The integral for the friction due to the water layer becomes elementary
\begin{equation} \label{B8}
F_{\rm fr} = \frac{\eta w V}{A} \int^l_0 
\frac{dx}{[\tanh ((l-x)/\lambda)]^{1/2}}= p_H w k^{1/2} \lambda^{3/4}
[l + l_a (0.5 \log(2)+ 0.25 \pi)].
\end{equation} 
As $k$ is proportional to $V$ the friction force vanishes as $V^{1/2}$.

\section{Heat Transfer in the water layer} \label{water}

The heat flow $J$ in the water layer is related to the temperature $T$ by the 
equation
\begin{equation} \label{D1} 
{\bf J} = - \kappa_{\rm w} \nabla T
\end{equation} 
In the stationary state the divergence of $J$ equals the heat source density, 
which is given by Eq.~(\ref{b3}) 
\begin{equation} \label{D2}
\kappa_{\rm w} \nabla^2 T = - \eta \frac{V^2}{h(x)^2}.
\end{equation} 
The solution of this equation has to be supplemented by the boundary 
conditions at the skate side $T=T_{\rm sk}$ and the ice side $T=T_{\rm ice}$. 
The main variation is parabolic in the downward $z$ direction.
In terms of the coordinate $z'$ with respect to the center of the layer
\begin{equation} \label{D3}
z'=z-d(x)-h(x)/2,
\end{equation}
we get the solution
\begin{equation} \label{D4}
T(z') = a + b z' - c z'^2.
\end{equation} 
The constant $c$ follows from Eq.~(\ref{D2}) as
\begin{equation} \label{D5}
c = \frac{\eta}{2 \kappa_{\rm w}} \, \frac{V^2}{h(x)^2}.
\end{equation} 
The boundary conditions give the values of the constants $a$ and $b$.
\begin{equation} \label{D6}
\left\{ \begin{array}{rcl}
T_{\rm sk} & = & a - b \, h(x)/2 - c [h(x)/2]^2, \\*[3mm]
T_{\rm ice} & = & a + b \, h(x)/2 - c [h(x)/2]^2. 
\end{array} \right.
\end{equation} 
With $\Delta T =T_{\rm ice}-T_{\rm sk}$ we find for $b$ 
\begin{equation} \label{D7}
b = \frac {\Delta T}{h(x)}.
\end{equation}  
The unimportant parameter $a$ follows by using this value in one of the 
Eqs.~(\ref{D6}).

With the temperature profile given we can determine the heat flows 
towards the skate and the ice. At the skate side we have a flow out of
the water layer
\begin{equation} \label{D8}
J_{\rm w-sk} = k_{\rm w} [b +  c \,h(x)]=\frac{k_w}{h(x)} [\Delta T + \Delta T_V],
\end{equation}
where $\Delta T_V$ is a temperature difference depending only on the
velocity $V$ and given by
\begin{equation} \label{D9}
\Delta T_V = \frac{\eta}{2 \kappa_w} \, V^2 = 
1.47 * 10^{-3}  \, V^2. 
\end{equation}
(With $V$ the numerical value in m/s and $\Delta T_V$  in centigrade.)
Likewise we have for the  flow towards the ice the value
\begin{equation} \label{D10}
J_{\rm w-ice} = -\kappa_{\rm w} [b -  c \,h(x)]= -\frac{\kappa_w}{h(x)} 
[\Delta T - \Delta T_V].
\end{equation} 
The fraction $\zeta_{\rm w}$ of the 
total heat produced in the layer towards the ice, is given by
\begin{equation} \label{D11}
\zeta_{\rm w} = \frac{J_{\rm w-ice}}{J_{\rm w-sk}} 
= \frac{1}{2} \left(1 - \frac{\Delta T}{\Delta T_v} \right).
\end{equation} 
The temperature at the ice side equals the melting temperature at the
pressure in the water layer.
At the skate side the temperature may be higher than this melting 
temperature, but cannot be lower. So $\Delta T < 0 \, $ and
the fraction will always be higher than or equal to 1/2. If 
$\, - \Delta T > \Delta T_v$, all heat flows towards the ice.

We note that, due to the layer thickness $h(x)$ in the denominator of
Eq.~(\ref{D10}), the temperature gradient at the water-ice interface
is huge.

\section{Heat Flows in the ice}\label{heatice}

The temperature distribution in the ice is governed by the heat equation
\begin{equation} \label{C2}
\frac{\partial T}{\partial t} = \alpha_{\rm ice} \Delta T + 
\left(\frac{\partial T}{\partial t}\right)_{\rm forced}.
\end{equation} 
First we have to find the expression for the temperature forcing. Take 
a point $x$ in the ice at time $t=0$. This point has experienced for earlier
times $t$ a temperature raise $T_m(x-Vt)-T_s$ at the surface, which we locate
for convenience at $z=0$. For the gradient in 
the $z$ direction this means a $\delta(z)$ dependence. So we find for the
temperature forcing
\begin{equation} \label{C5}
\left(\frac{\partial^2 T(x,z,t)}{\partial z \partial t} \right)_{\rm forced}
= \frac{\partial [T_m(x-Vt) -T_s]}{\partial t} \delta (z) =
-V \frac{\partial T_m(x-Vt) }{\partial x} \delta (z). 
\end{equation} 
This holds for times in the past up to $t_0$  
\begin{equation} \label{C6}
t_0=-(l-x)/V,
\end{equation} 
with $l$ the contact length. Let us first discuss the case where the pressure
at the tip has a melting temperature $T_m(l)$ {\it below} the surface 
temperature. Then we have to solve the following equation in the time interval 
$t_0 < t <0$ 
\begin{equation} \label{C7}
\frac{\partial^2T(x,z,t)}{\partial z \partial t} = \alpha_{\rm ice} \Delta 
\frac{\partial T(x,z,t)}{\partial z} - 
2 V \frac{\partial T_m(x-Vt) }{\partial x} \delta (z). 
\end{equation} 
We have inserted a factor 2 in the source term as it is easier to
solve the equation in the complete space $-\infty < z < \infty$ and to use
the symmetry between the upper and lower half $z$-plane. The differentiations
in the Laplacian $\Delta$ may be restricted to those in the $z$ direction,
since the variation in the $z$ direction is much larger than in the $x$ 
direction. The solution follows by Fourier transform in the $z$ direction
\begin{equation} \label{C8}
R_k (x,t) = \int^\infty_{-\infty} dz  
\frac{\partial T(x,z,t)}{\partial z}\, {\rm e}^{i k z} .
\end{equation} 
The equation for $R_k (t)$ reads
\begin{equation} \label{C9}
\frac{\partial R_k (x,t)}{\partial t} = - \alpha_{\rm ice} k^2 R_k (x,t) -
2 V \frac{\partial T_m(x-Vt) }{\partial x},
\end{equation} 
with the solution
\begin{equation} \label{C10}
R_k (x,t) = - 2 V \int^0_{t_0} dt' 
\frac{\partial T_m(x-Vt') }{\partial x}\, {\rm e}^{\alpha_{\rm ice} k^2 t'}.
\end{equation} 
The inverse Fourier transformation yields for the gradient at $z=0$
\begin{equation} \label{C11}
\left(\frac{\partial T(x,z,t)}{\partial z}\right)_{z=0} = 
- V \int^0_{t_0} dt' \frac{1}
{\sqrt{-\pi \alpha_{\rm ice} t'}} \frac{\partial T_m(x-Vt') }{\partial x}.
\end{equation}
Then changing the integration variable $t'$ to $x'=x-Vt'$ gives
\begin{equation} \label{C12}
\left(\frac{\partial T(x,z,t)}{\partial z}\right)_{z=0}  = 
-\left( \frac{V}{\pi \alpha_{\rm ice}} \right)^{1/2}
\int_x^l dx' \frac{1}{\sqrt{x'-x}} \frac{\partial T_m(x') }{\partial x'}.
\end{equation} 

This expression holds for the case $T_m(l)<T_s$. 
In the other case, when $T_m(l)>T_s$, the ice temperature is suddenly raised
at the tip by the amount $T_m(l)-T_s$ and one has in addition to the 
integral the contribution from this jump (leading to a $\delta$ function 
in the integral)
\begin{equation} \label{C13}
\delta \left(\frac{\partial T(x,z,t)}{\partial z}\right)_{z=0} = 
\left( \frac{V}{\pi \alpha_{\rm ice}} \right)^{1/2} \frac{T_m(l)-T_s}{\sqrt{l-x}}.
\end{equation} 
The combination of the integral and the jump are given in Eq.~(\ref{k2}).
In \cite{lozowski} only this jump is taken into account with $T_m(l)=0$.

Here we give for completeness the change in the layer equation as due
to this jump, in order to show how the layer equation of
Lozowski and Szilder \cite{lozowski} results. The fraction of heat 
available for melting is then reduced by the factor
\begin{equation} \label{C14}
\zeta = 1- \frac{2 \kappa_{\rm ice} [T_m(l)-T_s] h(x)} 
{\eta V^{3/2} \sqrt{\pi \alpha_i (l-x)}}.
\end{equation} 
We find $\zeta(\bar{x})$ by scaling $h(x)$ and $l-x$
\begin{equation} \label{C15}
\zeta(\bar{x}) = 1- q \frac{\bar{h} (\bar{x})}{[\bar{l} -\bar{x}]^{1/2}},
\end{equation} 
with the constant
\begin{equation} \label{C16}
q = \frac{\sqrt{2} \kappa_{\rm ice}}{\sqrt{\eta \pi \alpha_{\rm ice} \rho L_h}} 
\frac{T_m(l)-T_s}{V}= 1.825 \frac{T_m(l)-T_s}{V}.
\end{equation}  

The correction due to $\zeta$ changes the scaled equation (\ref{e7}) to
\begin{equation} \label{C17}
-\frac{d \bar{h}}{d \bar{x}} = \frac{1}{\bar{h} (\bar{x})}-
\frac{q}{\sqrt{\bar{l}-\bar{x}}}- 
\frac{\bar{x} + c_1}{c_2+ \bar{h}^3 (\bar{x})} \bar{h}^3(\bar{x}),
\end{equation} 
Eq.~(\ref{C17}) is the 
scaled version of a similar equation for the layer given in \cite{lozowski}.

It is interesting that $\zeta$ in Eq.~(\ref{C15}) approaches at the 
tip a finite value $\zeta(\bar{l})$, 
since $\bar{h}(\bar{l}-\bar{x})$ vanishes in the 
same way as the square root 
\begin{equation} \label{C18}
\bar{h}(\bar{l}-\bar{x}) \simeq a \sqrt{\bar{l} -\bar{x}}.
\end{equation} 
The amplitude $a$ satisfies the equation
\begin{equation} \label{C19}
\frac{a}{2} = \frac{1}{a} - q,\quad \quad {\rm or} \quad \quad 
a = \sqrt{2 + q^2}- q.
\end{equation} 
For $ q \rightarrow 0$ the amplitude $a=\sqrt{2}$ (as before in 
Eq.~(\ref{c4})) and for $q$ large,
the amplitude vanishes as $1/q$. So for low temperatures and slow velocities
the value of $a$ rapidly decreases, rendering the thickness of the 
water layer too thin to treat the layer as a hydrodynamic system.


\begin{thebibliography}{999}
\bibitem{bowden} F.~P.~Bowden and T.~P.~Hughes, Proc. R. Soc. A {\bf 2172} 
(1939) 280-298.\\
F.~P.~Bowden,  Proc. R. Soc. A {\bf 271} (1953), 462-478.  

\bibitem{schenau} J.~J. de Koning, G.~de Groot and G.~J. van Ingen Schenau,
J. Biomechanics {\bf 25} (1992) 565-571. 

\bibitem{faraday} M. Faraday, {\it Experimental Researches in Chemistry 
and Physics}, Taylor and Francies, London (1859) p. 372.  

\bibitem{rosenberg} B. Rosenberg, Physics Today (2005) Dec. 50-54.

\bibitem{amsterdam} B. Weber, Y. Nagata, S. Ketzetzi, F. Tang, W. J. Smit, H. J. Bakker, E.H.G. Backus, M. Bonn and D. Bonn. ``Molecular insight into the 
slipperiness of ice.'' Under review.\\
 B. Weber, PhD. thesis (2017) Univ. Amsterdam.

\bibitem{persson} B.~N.~J.~Persson, J. Chem. Phys. {\bf 143} (2015) 224701.\\
dx.doi.org/10.1063/1.4936299

\bibitem{leiden} T.~H.~A. van der Reep, Masters Thesis, (2014) Univ. Leiden.

\bibitem{lozowski} E.~P.~Lozowski and K.~Szilder, Int. Journ. of Offshore and
Polar Engineering {\bf 23} (2013) 04.

\bibitem{pomeau} M.~Le Berre and Y.~Pomeau, Int. Journ. of Non-linear Mech.
{\bf 75} (2015) 77-86. 

\bibitem{pourier} L.~Pourier, R.~I.~Thompson, E.~P.~Lozowski, S.~Maw and
D.~J.~Stefanyshyn, {\it 21st Int. Offshore and Polar Eng. Conf.} (2011) 
Maui, ISOPE, {\bf 3} 1071.

\bibitem{new} A.~Penny, E.~P.~Lozowski, T.~Forest, C.~Fong, C.~Maw, 
P.~Montgomery and N.~Sinha in {\it Physics and Chemistry of Ice} (2007)
495, W.~F.~Kuhn, editor, Roy. Soc. Chem.

\bibitem{nye} J.~F.~Nye Proc. Roy. Soc. A{\bf 219} 4 (1953) 477-489.

\bibitem{barnes} P.~Barnes, D.~Tabor and J.~C.~F.~Walter, Proc. R. Soc. Londen
A {\bf 324} (1971) 127-155.

\bibitem{karna} T.~Karna, Annals of Glacialogy {\bf 19} (1994) 114-120. 


\end{thebibliography}
\end{document}